# Structural and magnetic anomalies among the spin-chain compounds, $Ca_3Co_{1+x}Ir_{1-x}O_6$

S. RAYAPROL, KAUSIK SENGUPTA, and E.V. SAMPATHKUMARAN*
Tata Institute of Fundamental Research, Homi Bhabha Road, Colaba, Mumbai-400005, India.

***Abstract***

The results of x-ray diffraction, and ac and dc magnetization as a function of temperature are reported for a new class of spin-chain oxides, $Ca_3Co_{1+x}Ir_{1-x}O_6$. While the x= 0.0, 0.3, 0.5 and 1.0 are found to form in the $K_4CdCl_6$-derived rhombhohedral (space group $R\bar{3}c$ ) structure, the x= 0.7 composition is found to undergo a monoclinic distortion in contrast to a literature report. Apparently, the change in the crystal symmetry with x manifests itself as a change in the sign of paramagnetic Curie temperature for this composition as though magnetic coupling sensitively depends on such crystallographic distortions. All the compositions exhibit spin-glass anomalies with an unusually *large* frequency dependence of the peak temperature in ac susceptibility in a temperature range below 50 K , interestingly obeying Voger-Fulcher relationship even for the stoichiometric compounds.

## 1. Introduction

The investigation of quasi-one-dimensional magnetic compounds has picked up momentum in recent years. In this respect, the compounds of the type (Sr, Ca)$_3$ABO$_6$ (A, B = a metallic ion, magnetic or non-magnetic), crystallizing in the $K_4CdCl_6$-derived structures have started attracting the attention of physicists (see 1-16 and references therein), considering a variety of substitutions possible at A and B sites, thereby providing an ideal opportunity to tune relative strengths of interchain and intrachain magnetic interaction. The structure is characterized by the presence of chains of A and B ions running along c-direction arranged hexagonally forming a triangular lattice. We have found many magnetic anomalies in this class of compounds[8-16].

Here, we focus our attention on the series, $Ca_3Co_{1+x}Ir_{1-x}O_6$, the end members ($Ca_3Co_2O_6$ and $Ca_3CoIrO_6$, rhombhohedral structure, space group $R\bar{3}c$) of which have been reported to exhibit exotic magnetic properties. For instance, the compound, $Ca_3Co_2O_6$, has been proposed to serve as a rare example for "partially disordered antiferromagnetic structure (PDA)": That is, for an intermediate temperature (T) range (12-24 K), 2/3 of the ferromagnetic Co chains are antiferromagnetically coupled to each other, whereas the rest remain incoherent; as the T is lowered below 12 K, these incoherent chains have been proposed to undergo spin freezing, with the application of a magnetic field (H) inducing ferrimagnetic structure. There are several plateaus in the isothermal magnetization (M) data[3,6] in the magnetically ordered state, the features being dependent on temperature/field cycling-history, which are difficult to understand. On the other hand, the compound, $Ca_3CoIrO_6$, exhibits magnetic frustration effects around 30 - 50 K in the ac and dc M data, but without getting influenced by the application of magnetic fields as high as even 40 kOe, however without showing PDA structural features[14]. A common feature between these two compounds is that the ac magnetic susceptibility ($\chi$) exhibits an unusually large frequency ($\nu$) dependence in the vicinity of magnetic ordering temperature, uncharacteristic of spin glasses. It thus appears that these compounds are unconventional spin-glasses, that too in a stoichiometric environment due to topological frustration. It is of interest to investigate the magnetic behavior of the solid solution based on both the compounds, viz., on $Ca_3Co_{1+x}Ir_{1-x}O_6$, to see how the properties evolve from one end to the other. Kageyama et al[1] actually have investigated the dc $\chi$ behavior of this solid solution (x= 0.0, 0.2, 0.3, 0.5, 0.7, 0.8, 0.9, and 1.0) and found that the paramagnetic Curie temperature ($\theta_p$) changes sign around x= 0.7. This surprising finding needs confirmation and, if found to be true, there is a need to understand its origin. With this primary motivation and also to get a better insight into the magnetic behavior of this solid solution, we have carefully investigated ac and dc magnetization behavior of this solid solution as a function of temperature along with x-ray diffraction, the results of which are reported in this article.

* Email: sampath@tifr.res.in

## 2. Experimental

The compositions, x= 0.0, 0.3, 0.5, 0.7 and 1.0, were prepared by solid-state route as discussed in Ref. 1. The stoichiometric amounts of $CaCO_3$, $Co_3O_4$ and Ir (purity of all being more than 99.9%) were thoroughly mixed in an agate mortar. The pressed pellets were then calcined in air at 1173 K for a day. Then the specimens were ground again and the pressed pellets were heated in air for more number of days with intermediate grindings as follows: For all x except x= 1.0, 1323 K for 30 hrs; 1373 K for 30 hrs; 1423 K for 30 hrs. A sintering at a lower temperature of 1273 K (2 x 24 hrs) was required for x= 1.0, as otherwise the sample has been reported[1] to decompose. The samples were characterized by x-ray diffraction (Cu $K_\alpha$). The dc and ac magnetization measurements were performed employing a commercial superconducting quantum interference device (Quantum Design).

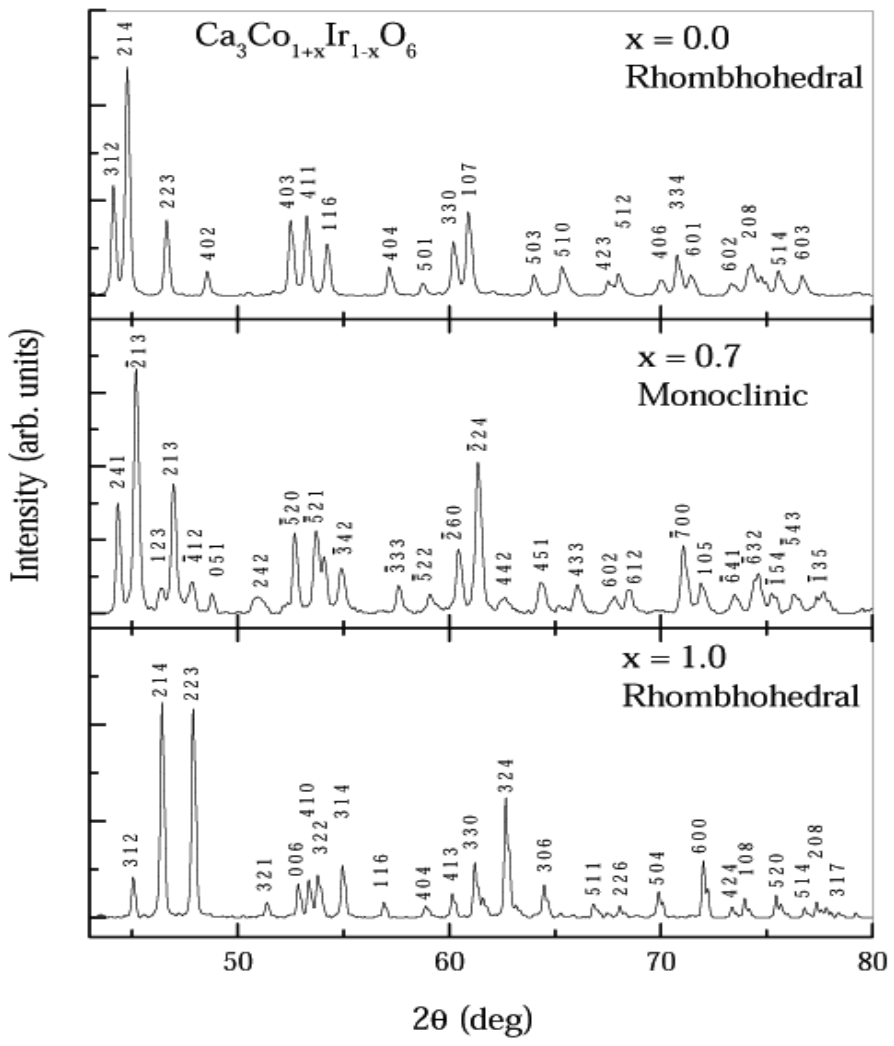

**Figure 1:** *X-ray diffraction pattern (Cu $K_\alpha$) at 300 K for the compositions x= 0.0, 0.7 and 1.0 of the series $Ca_3Co_{1+x}Ir_{1-x}O_6$ at higher angles to highlight the appearance of extra peaks due to monoclinic distortion for x= 0.7. The Miller indices are given.*

## 3. Results and discussion

### 3.1 X-ray diffraction

We first discuss the results of x-ray diffraction. We find that the patterns are essentially the same for all the compositions except for x= 0.7 and the diffraction patterns could be indexed to rhombhohedral structure ($R\bar{3}c$), as known earlier[1]. However, for x= 0.7, additional lines (see Fig. 1) are found to appear and entire pattern for this composition could be indexed only if one assumes a monoclinic distortion belonging to a space group of P2. Thus, the present results establish that there is a distortion of the crystal structure for an intermediate range. In order to demonstrate this point, the patterns for three compositions in the relevant range of angle are shown in Fig. 1 and the crystallographic parameters are listed in table 1.

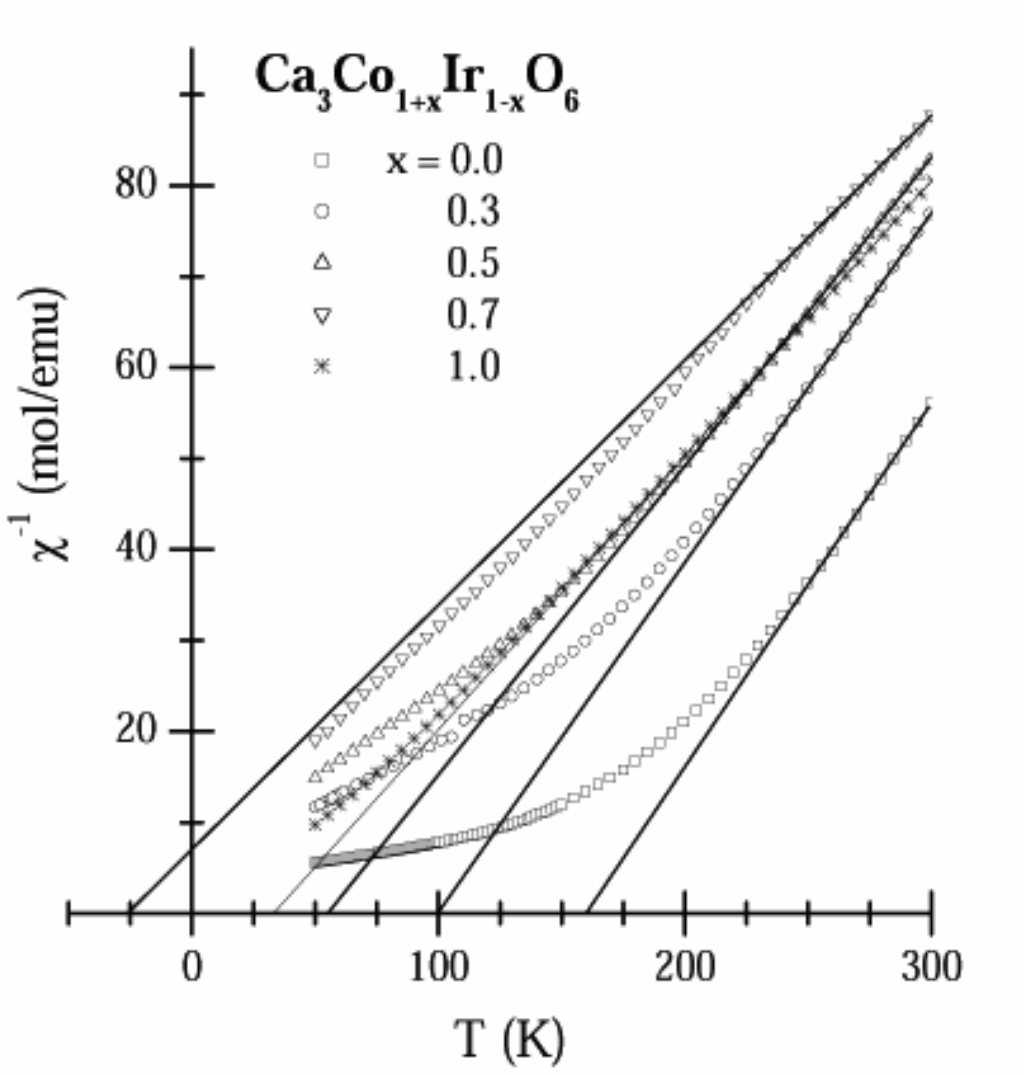

**Figure 2:** *Inverse dc susceptibility (H= 5 kOe) as a function of temperature above 50 K in the series, $Ca_3Co_{1+x}Ir_{1-x}O_6$. A line is drawn through the linear region above 250 K.*

### 3.2 Dc magnetization

The results of dc magnetization measurements are shown in Figs. 2-4. As known in the literature, the features in the low temperature $\chi$ data are complicated due to the onset of magnetic ordering. In order to see the trends in the values of $\theta_p$, we have shown the $\chi$ data in the paramagnetic state (above 50 K) in Fig. 2 in the form of inverse $\chi$ versus T. It is distinctly clear that the plots are not linear for any of the compositions. If one looks at the linear region in a narrow high T range, say 250 - 300 K, one can distinctly see that the sign of $\theta_p$ is positive for all compositions except for x= 0.7, confirming the findings of the previous report[1]. Viewed together with the inferences from the x-ray diffraction patterns discussed above, it is obvious that there is a close correspondence between crystal structure and magnetic properties, in the sense that monoclinic deformation modifies the sign of exchange interaction as well. In table 1, we have also listed the values of the magnetic moment ($\mu_{eff}$) obtained from the linear region. These values are in fair agreement with those reported in Ref. 1, though we observe marginally higher values at the $Ca_3CoIrO_6$-end. This discrepancy could be attributed to the uncertainties arising due to quasi-one-dimensional nature of the materials and, for the same reason, it is generally difficult to separate out the contributions from Co and Ir.

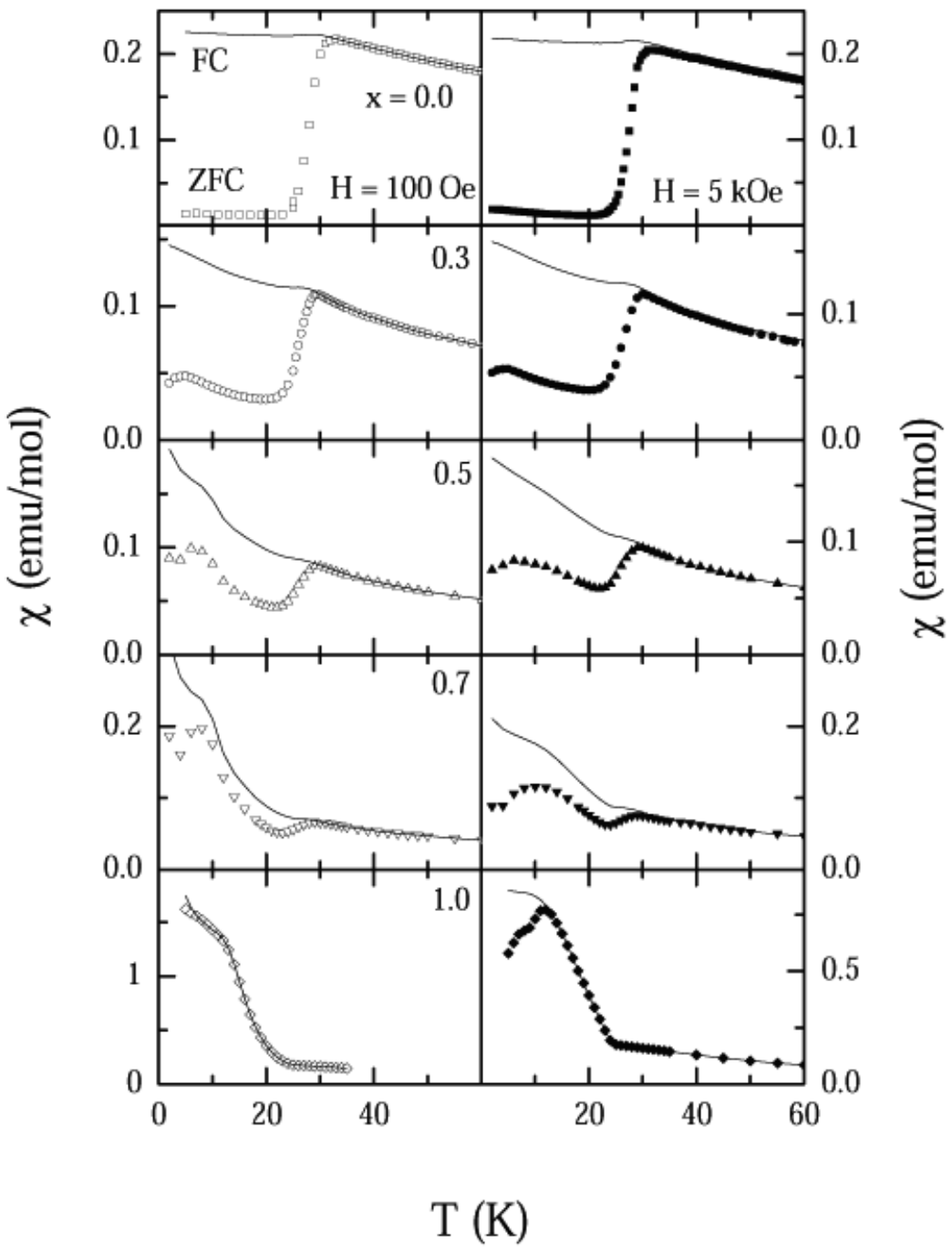

**Figure 3:** *Dc magnetic susceptibility as a function of temperature taken in the presence of 100 Oe and 5 kOe for $Ca_3Co_{1+x}Ir_{1-x}O_6$ for the zero-field-cooled and field-cooled states of the specimens.*

In order to understand how the properties are modified with x at low temperatures, we have tracked the dc $\chi$ behavior for the zero-field-cooled (ZFC) and field-cooled (FC) states of the specimens for H= 5 kOe and 100 Oe (see Fig. 3). We first look at the data for x= 1.0. There is a sudden increase in $\chi$ below 24 K, as though there is a ferromagnetic transition. In the ZFC-data for H= 5 kOe, as the T is lowered, this feature is followed by a peak at 12 K indicating the existence of another (antiferromagnetic-like) transition at this temperature. The ZFC-FC $\chi$ curves bifurcate below 12 K only, but not at 24 K, indicating that the 24 K transition is not of a spin-glass-type. All these observations are in broad agreement with those known in the literature. The new finding made here is that the bifurcation of ZFC-FC curves obtained at H= 100 Oe does not begin at 12 K, but near 7 K - the same T at which the ZFC-curve recorded at 5 kOe shows a hump, as though there is another magnetic transition around this temperature. These results imply that the magnetism of this compound is much more complex than what is believed in the literature. Now looking at the other end of the series, that is, for x= 0.0, ZFC-FC curves bifurcate around 30 K (as reported in Ref. 14) for both the field values. It is interesting to see that this bifurcation at 30 K is seen even for a small replacement of Co by Ir, say for x= 0.7, essentially superimposed over the features seen for $Ca_3Co_2O_6$ below 24 K. This tendency persists even for x= 0.3.

### *3.3 Ac magnetic susceptibility*

We have also carefully tracked the real part of ac $\chi$ behavior for all the compositions. Needless to emphasize that the two end members show a large $\nu$–dependence of $\chi$ in the vicinity of magnetic transitions (Fig. 4), as brought out in earlier articles[6, 14]. It is intriguing to note that the magnitude of the shift of the peak temperature ($T_f$) with increasing $\nu$ is uncharacteristic of conventional spin-glasses. In order to explore whether the spin-glass concepts as applied to dilute systems can still be applicable to these stoichiometric systems as well, we analysed the data in terms of Voger-Fulcher formula[17]:

$$\nu = \nu_o \exp[-E_a/k_B(T_f-T_a)]$$

where $E_a$ is the activation energy and $T_a$ is believed to be measure of intercluster interaction strength. That means a plot of $T_f$ versus $1/\ln(\nu_0/\nu)$ should be a straight line, which is interestingly found to be the case (see Fig. 5) if one assumes the value of $\nu_0$ to be the same ($10^8$ Hz) as that of ideal spin-glasses[17].

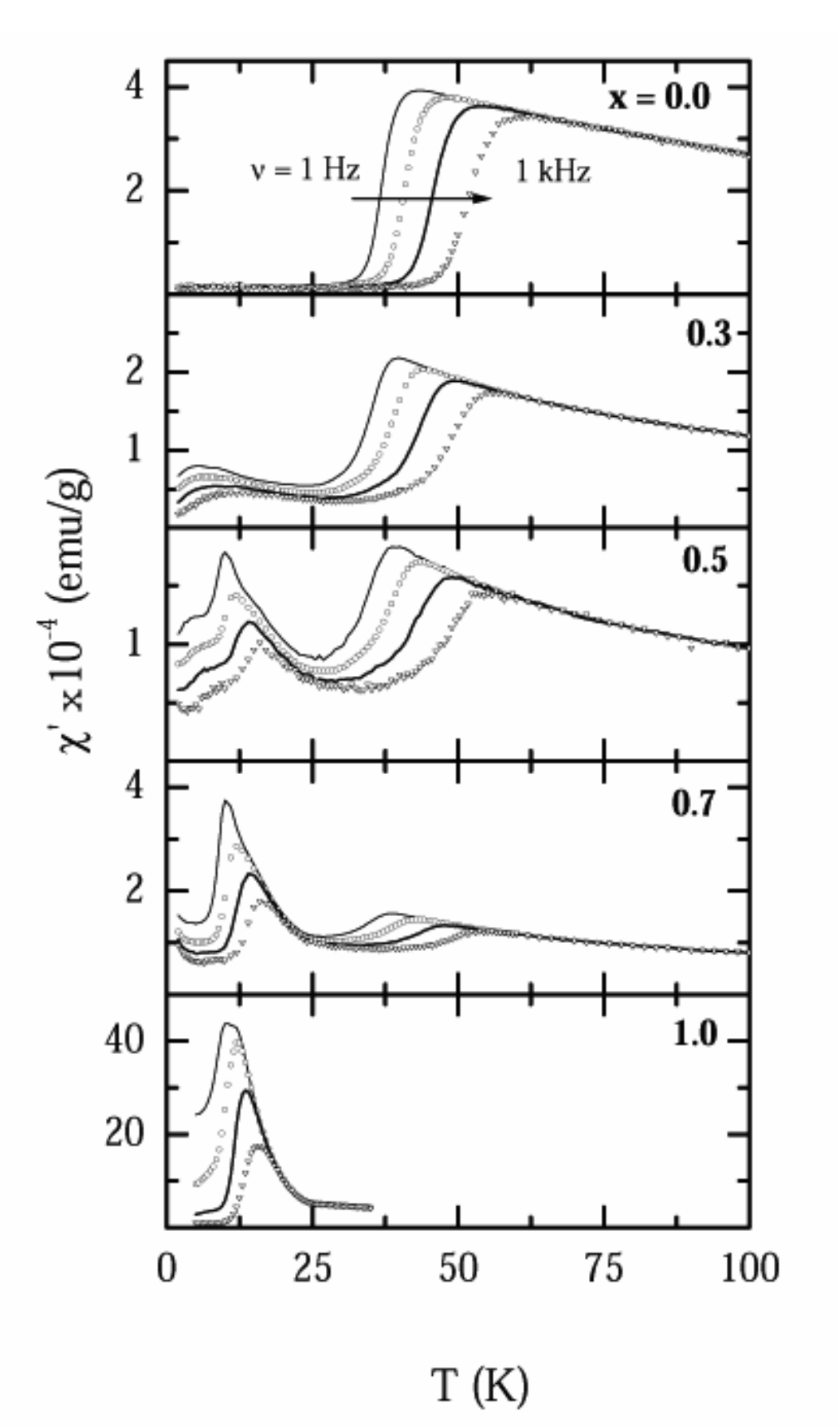

**Figure 4:** *Ac magnetic susceptibility (real part) as a function of temperature for $Ca_3Co_{1+x}Ir_{1-x}O_6$ recorded at various frequencies. The frequency (1, 10, 100, 1000 Hz) increases with the peak temperatures in the direction of the arrow.*

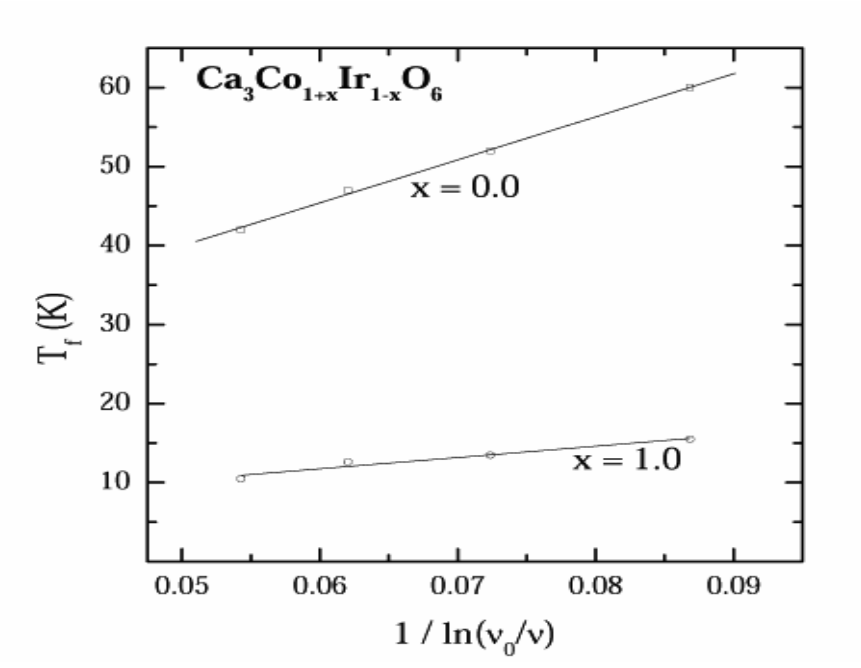

**Figure 5:** *The plot of peak temperature in the real part of ac susceptibility versus $1/\ln(\nu_0/\nu)$ for $Ca_3Co_2O_6$ and $Ca_3CoIrO_6$ to test the validity of Voger-Fulcher relationship.*

Thus, these compounds turned out to be novel examples for spin glass behavior in the stoichiometric situation. Now turning to solid solutions, it is to be noted that the observed ac $\chi(T)$ above 10 K for every composition is a superimposition of the features observed for the end members (similar to the dc $\chi$ behavior) with the peak temperature being the same for all compositions for a given $\nu$ (that means, there are two prominent peak temperatures for intermediate compositions). Therefore one naturally tends to doubt whether the intermediate compositions are physical mixtures of two end compositions. A careful look at all the data reported in the article convincingly reveals that this is not the case: (i) The change in crystal symmetry and the sign of $\theta_p$ for x= 0.7 can not be understood in terms of a physical mixture of the two end compounds; (ii) the 7K-feature in the ac $\chi$ data is more prominent for the intermediate compositions, particularly for x= 0.5. (iii) The isothermal M(T) data recorded at 5 K (see Fig. 6) for intermediate compositions is not found to be a linear extrapolation of those of end members. Therefore, the observed ac and dc $\chi(T)$ behavior for intermediate compositions are quite fascinating, warranting further studies for better understanding.

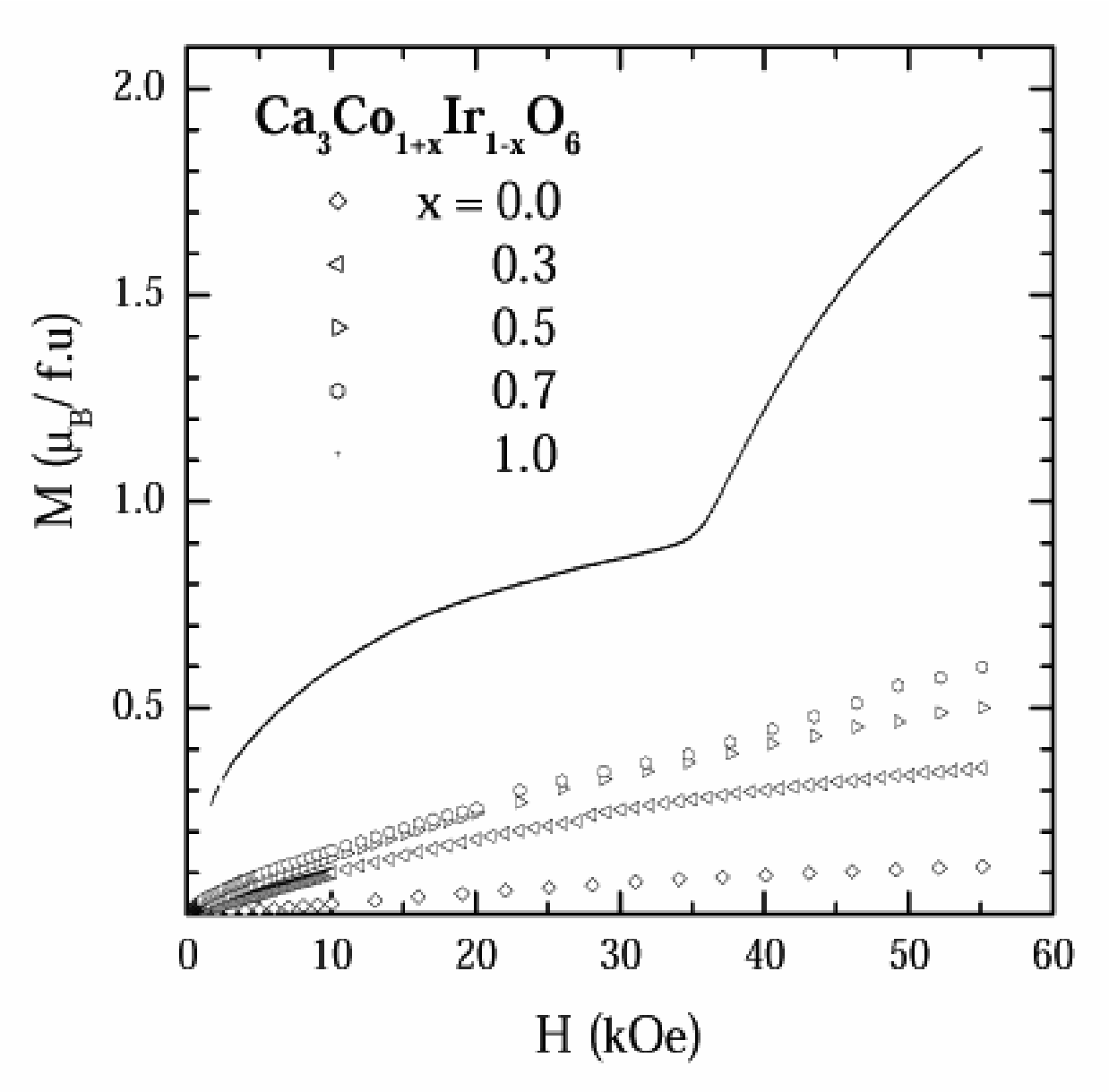

**Figure 6:** *Isothermal magnetization at 5 K for $Ca_3Co_{1+x}Ir_{1-x}O_6$ and the lines through data points serve as guides to the eyes.*

## 4. Conclusions

To conclude, we have reported ac and dc magnetization behavior of the quasi-one-dimensional magnetic compounds, $Ca_3Co_{1+x}Ir_{1-x}O_6$. It is interesting that all these compositions exhibit novel spin-glass characteristics, with the new point emphasized here being the validity of the Voger-Fulcher relationship in the ac $\chi$ data even in stoichiometric compounds. For some intermediate compositions, there appears to be a crystallographic distortion to monoclinic structure (as inferred from the x-ray diffraction patterns at room temperature) which is apparently responsible for the anomalies in $\theta_p$ reported in previous literature (which is also confirmed here). This implies that there is a correlation between structure and magnetism in the paramagnetic state of this class of compounds. It is of interest to carry out low temperature (below 50 K) crystallographic studies in order to explore the existence of similar correlations even in the vicinity of magnetic ordering temperature.

Acknowledgements:
We thank Kartik K Iyer for his experimental help during the course of this work.

**Table 1:**

Composition (x), crystal structure, space group, the lattice constants (a and c), the paramagnetic Curie temperature ($\theta_p$) and effective moment ($\mu_{eff}$) obtained from the T range 250-300 K and the temperature ($T_o$) at which the features due to magnetic transitions are seen as inferred in the susceptibility data, for the series $Ca_3Co_{1+x}Ir_{1-x}O_6$.

| x | Structure/ Space group | a (Å) | b (Å) | c (Å) | $\theta_p$ (K) ($\pm$ 2 K) | $\mu_{eff}$ ($\mu_B$) ($\pm$ 0.1 $\mu_B$) | $T_o$ (K) |
|---|---|---|---|---|---|---|---|
| 0.0 | Rhombhohedral R $\bar{3}c$ | 9.214 | 9.214 | 10.900 | 160 | 4.48 | 31 |
| 0.3 | Rhombhohedral R $\bar{3}c$ | 9.193 | 9.913 | 10.802 | 99 | 4.56 | 30, 7 |
| 0.5 | Rhombhohedral R $\bar{3}c$ | 9.188 | 9.188 | 10.768 | 55 | 4.85 | 30, 12, 7 |
| 0.7 | Monoclinic* P 2 | 9.270 | 9.721 | 6.662 | -26 | 5.45 | 29, 12, 7 |
| 1.0 | Rhombhohedral R $\bar{3}c$ | 9.078 | 9.078 | 10.384 | 33 | 5.14 | 24, 12, 7 |

* $\beta = 92.31^0$